\documentclass[a4paper]{article}

\usepackage{INTERSPEECH2019}
\usepackage{amsmath}
\usepackage{epstopdf}
\usepackage{booktabs}
\usepackage{color}
\usepackage{amssymb}

\usepackage{mathtools}
\usepackage{cite}
\usepackage[ruled]{algorithm2e}
\usepackage{graphicx}
\usepackage{subfigure}

\title{A Multi-Task Learning Framework for Overcoming the Catastrophic Forgetting in Automatic Speech Recognition}
\name{Jiabin Xue, Jiqing Han, Tieran Zheng, Xiang Gao and Jiaxing Guo}
\address{
  School of Computer Science and Technology, Harbin Institute of Technology, Harbin, China
  }
\email{\{xuejiabin, jqhan, zhengtieran, xianggao, guojiaxing\}@hit.edu.cn}
\begin{document}

\maketitle
\begin{abstract}
Recently, data-driven based Automatic Speech Recognition (ASR) systems have achieved state-of-the-art results.
And transfer learning is often used when those existing systems are adapted to the target domain, e.g., fine-tuning, retraining.
However, in the processes, the system parameters may well deviate too much from the previously learned parameters.
Thus, it is difficult for the system training process to learn knowledge from target domains meanwhile not forgetting knowledge from the previous learning process, which is called as catastrophic forgetting (CF).
In this paper, we attempt to solve the CF problem with the lifelong learning and propose a novel multi-task learning (MTL) training framework for ASR. It considers reserving original knowledge and learning new knowledge as two independent tasks, respectively.
On the one hand, we constrain the new parameters not to deviate too far from the original parameters and punish the new system when forgetting original knowledge.
On the other hand, we force the new system to solve new knowledge quickly. Then, a MTL mechanism is employed to get the balance between the two tasks.
We applied our method to an End2End ASR task and obtained the best performance in both target and original datasets.

\end{abstract}

\section{Introduction}
\label{sec:intro}
Deep neural network(DNN) is currently becoming a successful model in the automatic speech recognition (ASR) field with the continuous development of the theory and technology of deep learning \cite{sainath2015convolutional, DBLP:conf/icassp/ValtchevOWY96, amodei2016deep}.
However, the performance of them is heavily affected by the data scale and domain coverage of training data since they are data-driven.

Various new domains are constantly emerging with the wide application of the Internet.
Thus, it is necessary that the ASR systems adapt these domains as soon as possible so as to meet the application requirements.
Two transfer learning methods, i.e., retraining (RT) \cite{DBLP:conf/interspeech/PaulikFSSSW05} and fine-tuning (FT) \cite{yu2010roles}, are often used when the existing ASR systems are adapted to the target domain.
For the RT of the ASR system, it usually conducts with all training data both from the original and target domains.
This method, however, has some shortcomings, e.g., it cannot effectively model the ASR system for the target domain when the difference between the original and target datasets is large, it can achieve the best performance neither on the original nor the target domain.
For the FT of the ASR system, it performs with only the training data from the target domain, and the system parameters are initialized with the original ones.
This method can accelerate the domain adaptation process by sharing the learned parameter with the new system which is trained not from scratch.
However, the performance on the original domain will be considerably degraded although it achieves a better performance on the target domain.
In summary, both the above two methods are prone to the catastrophic forgetting (CF) \cite{french1999catastrophic} problem.

CF is a serious common problem in artificial intelligent systems.
It results in the failure of the aforementioned systems to learn the new knowledge fast without forgetting previously acquired knowledge.
One of the most commonly studied methods to overcome the CF is lifelong learning \cite{kirkpatrick2017overcoming,rusu2016progressive}.
However, the existing methods in lifelong learning cannot be applied to large-scale machine learning problems.
Therefore, we attempt to solve the domain adaptation problem of ASR by using improved lifelong learning in this paper, and to the best of our knowledge, it is the first attempt to employ the lifelong learning in this field.

The current mainstream lifelong learning methods can be divided into two approaches depending on whether the network architecture has been changed during the learning process.
The first one is to adopt a complex fixed network architecture with a large capacity.
And when training the network for consecutive tasks, some regularization term is enforced to avoid that the model parameters deviate too much from the previously learned parameters according to their significance to old tasks\cite{kirkpatrick2017overcoming}.
The second one is to dynamically extend the network structure to accommodate new tasks while the previous architecture parameters are kept unchanged, for instance, progressive networks \cite{rusu2016progressive} which extend the architecture with a fixed node or layer size, and dynamically expandable network (DEN) \cite{DBLP:conf/iclr/YoonYLH18} which introduces group sparse regularization when new parameters are added to the original networks.

At present, most methods based on the dynamic extended network cannot be effectively applied in large-scale speech recognition tasks, since the continuous increase of tasks has increased the complexity of the model structure.
This situation results in an increase in the training and reasoning times and failures to meet the requirements of real-time processing and computational resources.
Therefore, we select the first fixed model for the model training.

A surprising result in statistics is Stein몶s paradox \cite{stein1956inadmissibility}, which emphasizes that learning multiple tasks together can obtain superior performance than learning each task independently.
Thus, in order to overcome the CF problem, we  attempt to divide the task of domain adaptation into two parts: reserving original knowledge and learning new knowledge.
Then, a multi-task learning (MTL) \cite{caruana1997multitask} mechanism is explored to balance the two tasks.
Many studies have shown that different configurations of the system can get the same performance. Accordingly, we try to find the best configurations of the target domain near to the configurations on the original dataset.

In accordance with the above-mentioned analysis, we propose a novel MTL training framework for ASR to solve the CF problem.
This framework considers the reserving original knowledge and learning new knowledge as two independent tasks, and an MTL mechanism is employed to balance the two tasks.
Experimental results on the LibriSpeech corpus \cite{panayotov2015librispeech} and the Switchboard corpus \cite{godfrey1992switchboard} show that our proposed method can achieve a better performance than FT and RT on these two datasets.

\section{The Proposed Method}
\subsection{Reason Analysis of Catastrophic Forgetting}

Suppose we have a dataset $Data0$ of the original domain and have trained a model $Model_{\theta_{0}^{*}}$ on it.
We collect another dataset $Data1$ belonging to the target domain.
The problem is how to acquire a fresh $Model_{\theta_{1}^{*}}$ that has the preferable performance on the original domain and target domain simultaneously thereby expanding the covered domain of the DNN based ASR system.

A DNN contains multiple layers projection, which comprises a linear projection and followed by element-wise nonlinearities projection.
Learning a task consists of adjusting the set of weights of the linear projections to optimize the performance.
Many works have pointed out that different configurations of $\theta$ possibly can get the same performance \cite{hecht1992theory, sussmann1992uniqueness}.
Besides, the original dataset and the new one all belong to the ASR domain.
Thus, their optimal parameter space may well not deviate too far away and have an overlapping part in which we can find an optimal parameter which has a good performance on both the original domain and the target domain.
In this paper, we propose a new MTL ASR training framework to avoid the CF, i.e., Multiple Task Learning for Catastrophic Forgetting (MTLCF).
It tries to look for the optimal parameters of a new model near the parameters of the original model.

\begin{figure}
\centering
\includegraphics[width=1in]{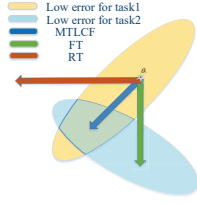}
\caption{Schematic parameter space of the original knowledge and the new knowledge.}
\label{fig:Figure1}
\end{figure}

In Figure \ref{fig:Figure1}, we illustrate training trajectories in a schematic parameter space with parameter regions leading to a good performance on original knowledge (yellow) and on new knowledge (blue).
After learning the first task, the parameters are at $\theta^{*}$.
From this figure, we can see that if we take gradient steps according to the FT alone (green arrow), we may only minimize the loss of new knowledge but destroy what we have learned for original knowledge.
Then, if we take gradient steps according to the MI alone (red arrow), the restriction imposed is too severe, and we can remember original knowledge only at the expense of not learning the new knowledge.
Our MTLCF, conversely, hopes to find a solution for the new knowledge without incurring a significant loss on the original knowledge (blue arrow).

\subsection{Multi-Task Learning for Catastrophic Forgetting}
\subsubsection{Multi-Task Object Function of MTLCF}

Our method splits the problem of overcoming the CF into two independent tasks, i.e., $task1$ and $task2$, and jointly optimizes the two tasks.
The $task1$ is used to avoid the new model leaving the optimal parameter space of the original domain, so that the parameter in this space can achieve a good performance in the original domain.
Further, the $task1$ is divided into two subtasks, i.e., $subtsak1$ and $subtask2$.

In the $subtask1$, we restrict the retrained model not to deviate too far from the original one by minimizing the KL divergence of the output distributions between the original model and the retrained model,
\begin{equation}
\begin{aligned}
\label{eq1}
subLoss1 &= T^{2}\sum_{i=1}^{m}KL(\mathbf{L}_{1}^{i}/T,\mathbf{L}_{0}^{i}/T),\\
         &= T^{2}\sum_{i=1}^{m}\mathbf{L}_{1}^{i}/T \cdot (log(\mathbf{L}_{1}^{i}/T)\\
         & - log(\mathbf{L}_{0}^{i}/T)),
\end{aligned}
\end{equation}
where $subLoss1$ is the loss of $subtask1$, $\mathbf{L}_{0}^{i}$ and $\mathbf{L}_{1}^{i}$ are the output distributions of the $Model_{\theta_{0}^{*}}$ and the $Model_{\theta_{1}}$ with a pair of sample $\{x_{0}^{i},y_{0}^{i}\}$ sampled from the $Data0$, $m$ is the batch size, $KL(\mathbf{L}_{1},\mathbf{L}_{0})$ is the Kullback-Leibler Divergence, $T$ is a sharpness parameter.
In this paper, we use the fixed $T=1$ to reduce the number of hyperparametrices.

Then, we consider to employ the Viterbi algorithm to find the optimal path in the Connectionist Temporal Classification (CTC) \cite{graves2006connectionist} loss, so even if the output distributions of the above two models are similar, it is also possible to find two different paths.
For the $subtask2$, its main work is to minimize the CTC Loss of the training data of the original domain.
From this subtask, we can calculate the degree to which the new model forgets the original knowledge and then punish it.
Thus, the loss of $task1$ is as follows,
\begin{equation}
\begin{aligned}
\label{eq1}
Loss1 &= \alpha subLoss1 + (1 - \alpha) subLoss2,\\
         &= \alpha (T^{2}\sum_{i=1}^{m}\mathbf{L}_{1}^{i}/T \cdot (log(\mathbf{L}_{1}^{i}/T) \\ &- log(\mathbf{L}_{0}^{i}/T)))\\ &+ (1 - \alpha) \sum_{i=1}^{m} CTCLoss(\mathbf{L}_{1}^{i}, y_{0}^{i}),
\end{aligned}
\end{equation}
where $Loss1$ is the loss of the $task1$, $\alpha$ is a hyperparametric, which controls the penalty for forgetting, $CTCLoss(x,y)$ is the loss function of CTC Loss between $x$ and $y$.

We force the new model to solve the new knowledge quickly by the $task2$, which optimize the CTC Loss of the retained model on the target domain training data.
Therefore, the loss in this method is as follows,
\begin{equation}
\begin{aligned}
\label{eq1}
Loss &= \beta Loss1 + (1 - \beta) Loss2,\\
         &= \beta(\alpha (T^{2}\sum_{i=1}^{m}\mathbf{L}_{1}^{i}/T \cdot (log(\mathbf{L}_{1}^{i}/T) \\ &- log(\mathbf{L}_{0}^{i}/T)))\\ &
         + (1 - \alpha) \sum_{i=1}^{m} CTCLoss(\mathbf{L}_{1}^{i}, y_{0}^{i}))\\ &
         + (1 - \beta) \sum_{j=1}^{m}CTCLoss(\mathbf{L}_{2}^{j}, y_{1}^{j}) ,
\end{aligned}
\end{equation}
where $Loss2$ is the loss of the $task2$, $\beta$ is a hyperparametric, which controls the speed to solve the new knowledge, $\mathbf{L}_{2}^{i}$ is the output distribution of the $Model_{\theta_{1}}$ with the pair of sample $\{x_{1}^{i},y_{1}^{i}\}$ sampled from the $Data1$.

\begin{figure}
\centering
\includegraphics[width=2.8in]{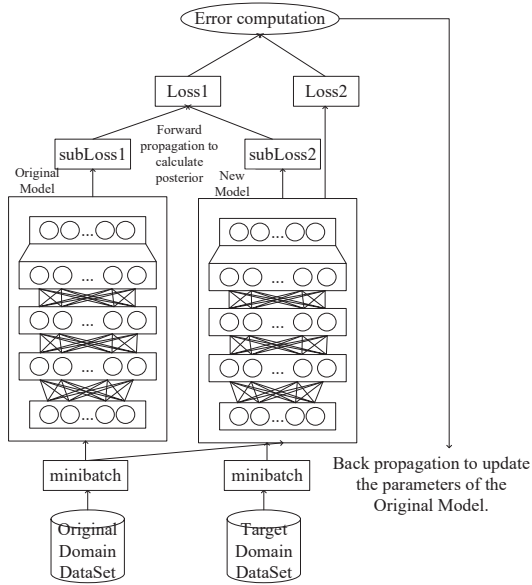}
\caption{
The structure of the MTLCF.
It uses two data set which are selected from two different data distributions.
The framework consists of an original model $Model_{\theta_{0}^{*}}$, which is trained in $Data1$ and a new model $Model_{\theta_{1}}$, which is copied from $Model_{\theta_{0}^{*}}$.
Only $\theta_{1}$ is updated during the process of back propagation.
}
\label{fig:Figure2}
\end{figure}

\subsubsection{The Structure of MTLCF}

We display the structure of MTLCF in Figure \ref{fig:Figure2}.
It can be seen that our MTLCF is composed of two task networks, which can be any network of interest in solving a particular task, such as Convolutional Neural Networks (CNN) \cite{chua1988cellular} and Long Short-Term Memory (LSTM) \cite{hochreiter1997long} Neural Networks. In this paper, we use the LSTM Deep Neural Networks (LDNN) \cite{sainath2015convolutional} architecture as the task network to do an End-to-End ASR task.

The MTLCF training procedure is described in Algorithm 1.
In the training process, we first copy a new model from the current model, and then fix the parameter of the original model and only train the parameter of the new model.

\begin{algorithm}
\caption{MTLCF}
\KwIn{$Data0$, the original domain data. $Data1$, the target domain data. $Model_{\theta^{*}}$, the original model. $m$, the batch size. }
\KwOut{$Model_{\theta_{1}^{*}}$, the target model}
Copy $Model_{\theta_{1}}$ from $Model_{\theta^{*}}$\;

\While{$\theta_{1}$ has not converged}{
모모Sample ${(x_{0}^{i},y_{0}^{i})\sim Data0}_{i=1}^{m}$ a batch of original speech data.\;
모모Sample ${(x_{1}^{i},y_{1}^{i})\sim Data1}_{i=1}^{m}$ a batch of new speech data.\;
    $L_{0}^{i} {\leftarrow} Model_{\theta_{0}}(x_{0}^{i})$\;
    $L_{1}^{i} {\leftarrow} Model_{\theta_{1}}(x_{0}^{i})$\;
    $L_{2}^{i} {\leftarrow} Model_{\theta_{1}}(x_{1}^{i})$\;
    $Loss {\leftarrow} \sum_{i=1}^{m} Loss(L_{0}^{i},L_{1}^{i},L_{2}^{i})$\;
    $g\theta {\leftarrow} \frac{\partial Loss}{\partial \theta_{1}}$\;
    $\theta_{1} {\leftarrow} \theta_{1} - Optimizer(\theta_{1}, g\theta)$\;
}
$\theta_{1}^{*} {\leftarrow} \theta_{1} $\;
\end{algorithm}

\section{Experimental Details}
\subsection{Experiment Data}
Our experiments are conducted on the LibriSpeech corpus, which consists of 1000h of training data, and the 300-hour Switchboard English conversational telephone speech task, both of them are the most studied ASR benchmarks today \cite{saon2017recent, xiong2017microsoft, vesely2013sequence, povey2016purely, medennikov2016improving}.
We select data from the train-clean-360 folder, the dev-clean and the test-clean folder for the training set, development set and test set for $Data1$, respectively.
For $Data2$, we select 95\% and 5\% from Switchboard-1 Release 2 (LDC97S62) as a training set and development set, respectively.
And then, we select data from the Hub5 2000 (LDC2002S09) evaluation set, which contains 20 ten-minute conversations from Switchboard (SW) and 20 ten-minute conversations from CallHome English (CH), as the test set.

The acoustic feature is 80-dimensional log-mel filterbank energies, computed using a 25ms window every 10ms, which is extracted using kaldi \cite{povey2011kaldi}.
Similar to the low frame rate (LFR) model \cite{pundak2016lower}, at the current frame $t$, these features are stacked with two frames to the left and downsampled at a 30ms frame rate, producing a 240-dimensional feature vector.

\subsection{Model Training}
For evaluation, all neural networks are trained using the CTC objective function with character targets and TensorFlow \cite{abadi2016tensorflow}.
Similar to the \cite{sainath2015convolutional}, all networks consist of a LDNN model, which comprises three bidirectional LSTM (Bi-LSTM) layers followed by a Rectified Linear Unit (ReLu) layer and a linear layer.
Each LSTM consists of $320$ cells. The fully connected ReLu layer has $1024$ nodes which are followed by a softmax layer with $46$ output units.

The stage of training is similar to \cite{audhkhasi2018building}.
The weights for all layers of $Model_{\theta_{0}^{*}}$ are uniformly initialized to lie between -0.05 and 0.05.
And the weights for all layers of $Model_{\theta_{1}}$ are copied from $Model_{\theta_{0}^{*}}$.
All networks are trained using Adam \cite{DBLP:journals/corr/KingmaB14} with a learning rate of $0.001$.
The learning rate is halved whenever the held-out loss does not decrease by at least 10\%.
We clip the gradients to lie in $(-5, 5)$ to stabilize training. The training data sorted by length.
\subsection{Experiment Results}

\subsubsection{Analysis of the Models Convergence Speed}

We first analyze the convergence rate of the models for RT, FT and MTLCF with $\alpha=\beta=0.5$ (MTLCF$\_0.5\_0.5$), and the results are shown in Figure 3.

\begin{figure}[!htbp]
\subfigure[]{
\label{fig:figure3:a}
\begin{minipage}[t]{0.5\linewidth}
\includegraphics[width=1.6in]{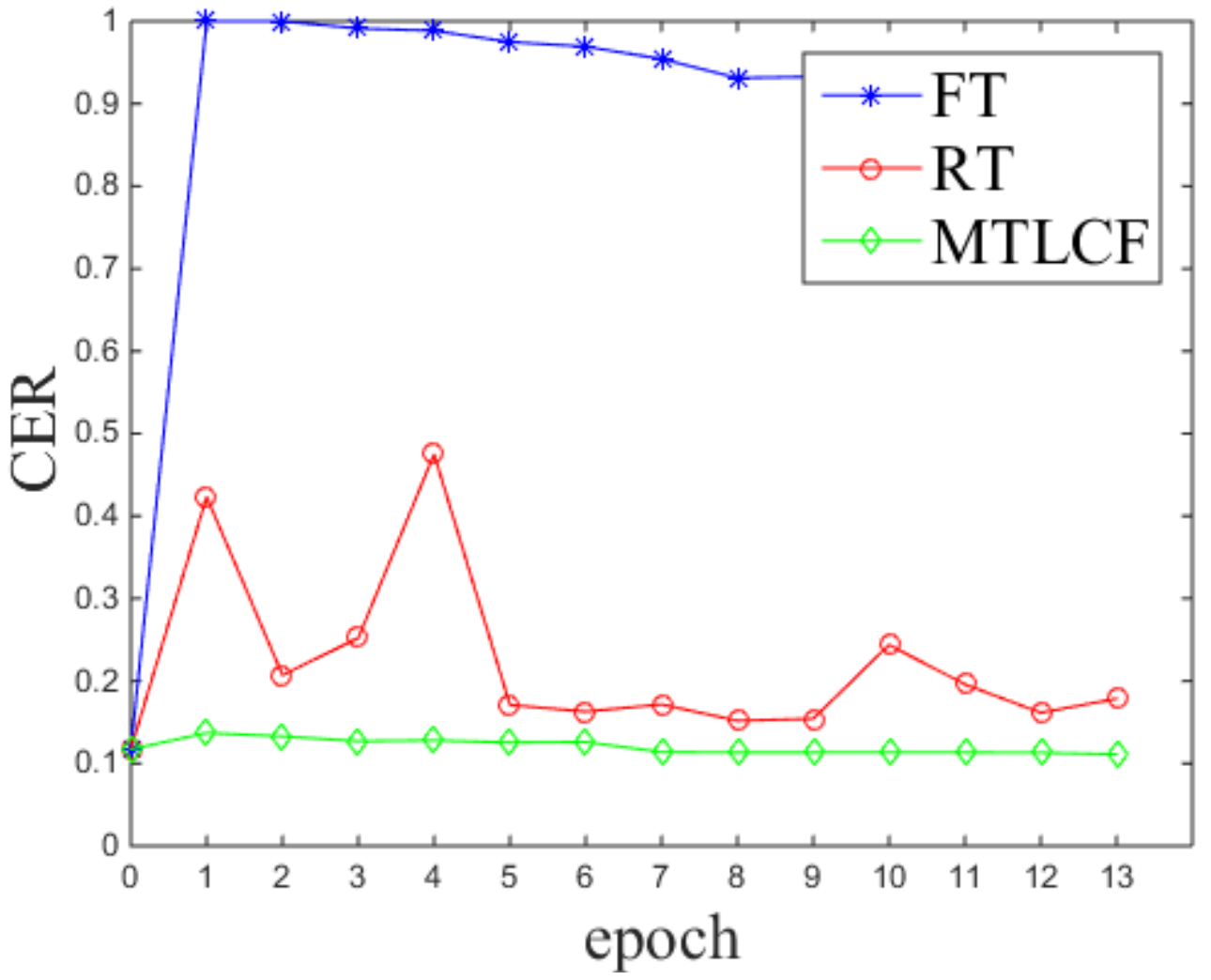}
\end{minipage}
}%
\subfigure[]{
\label{fig:figure3:b}
\begin{minipage}[t]{0.5\linewidth}

\includegraphics[width=1.6in]{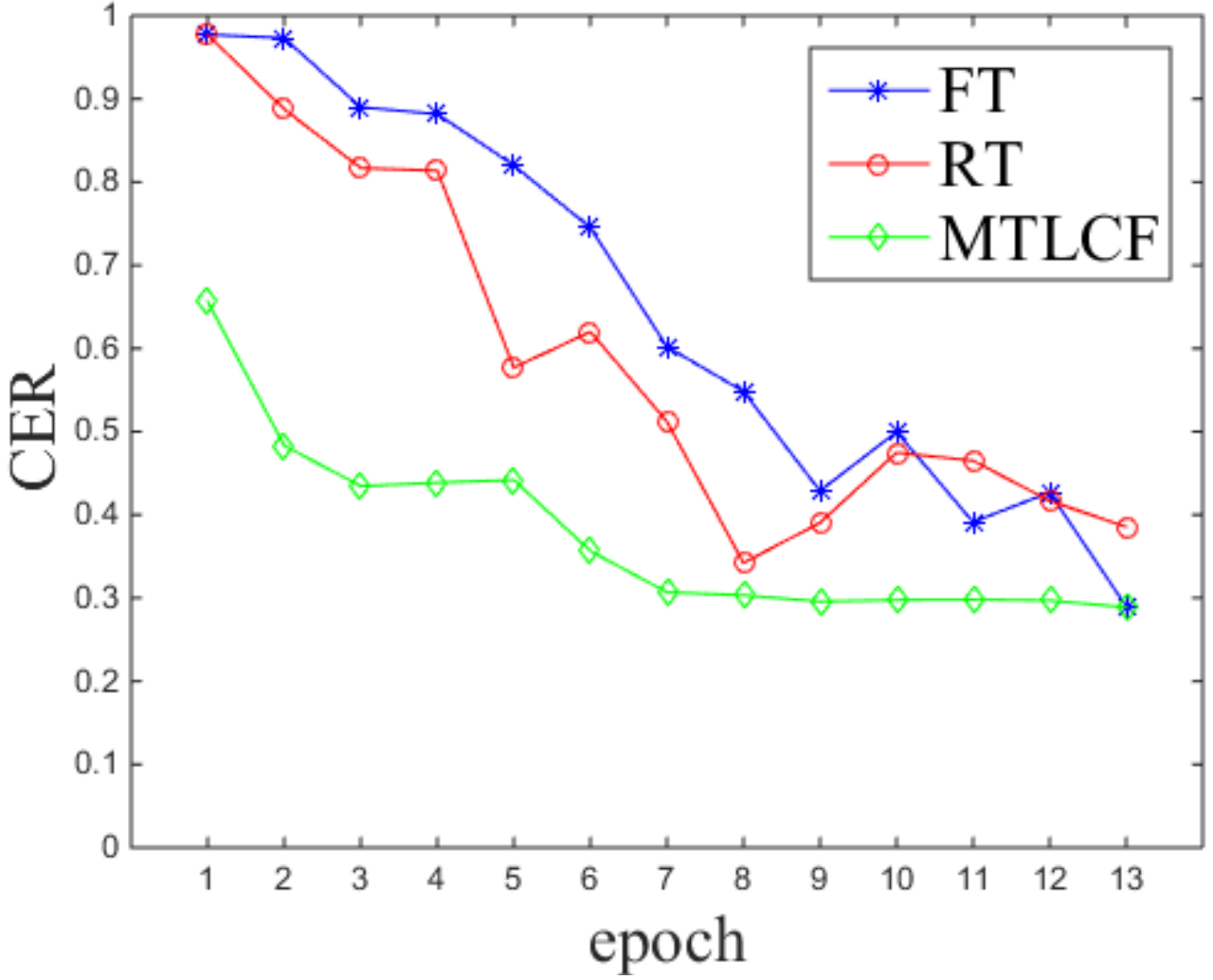}
\end{minipage}
}%

\caption{The description of the converge rate of $RT$, $FT$ and $MTLCF\_0.5\_0.5$ on the $test_{org}$ (a) and on the $test_{tar}$ (b).}
\label{Figure3}
\end{figure}

In Figure \ref{Figure3}\subref{fig:figure3:a}, we show the convergence rate of these models on the original domain test set ($test_{org}$), where the point in the epoch $0$ presents the character error rate (CER) of the original model on the $test_{org}$.
As can be seen from Figure \ref{Figure3}\subref{fig:figure3:a}, the FT forgets all the learned knowledge at the end of the first epoch, since it suffers a serious CF.
The learning of RT is unstable, it often forgets a lot of learned knowledge and then learn them at once. However, it cannot achieve the initial performance.
The MTLCF$\_0.5\_0.5$ is stable on the $test_{org}$ without forgetting the previous knowledge.
And the convergence rate of these models on the target domain test set ($test_{tar}$) is shown in Figure \ref{Figure3}\subref{fig:figure3:b}.
We can see that the fastest converge rate is achieved by the MTLCF$\_0.5\_0.5$, which is converged within $7$ epochs.

From  the above analysis, it can be seen that the proposed model achieves the best convergence speed on both the original and target domain.

\subsubsection{Analysis of the CER}

In Table \ref{Table1}, we give the performance of the $Model_{\theta_{0}^{*}}$, which is trained on the original domain training data using a random initialization, on the $test_{org}$ and $test_{tar}$.
$Model_{\theta_{0}^{*}}$ achieves a CER of $0.117$ on the $test_{org}$, but the CER on the $test_{tar}$ reaches $1.05$.
Therefore, we can find that the data distribution of the original domain and the target domain is quite different.

\begin{table}[!htbp]
\centering
\caption{Performance of the $Model_{\theta_{0}^{*}}$ on the $test_{org}$ and $test_{tar}$}
\begin{tabular}{ c | c | c }
\hline
 Training data set  &   $test_{org}$(CER) & $test_{tar}$(CER)\\ \hline

 LibriSpeech    & $0.117302$  & $1.05$
\\\hline
\end{tabular}
\label{Table1}
\end{table}

We show the final convergence results of those models in Table \ref{Table2}.
On the one hand, we analyze the results of an equal amount of training data.
It can be seen from the top three rows in Table \ref{Table2} that the FT basically forgets the learned knowledge, but it can achieve good performance on the $test_{tar}$.
The RT model can reach the optimal convergence result on neither $test_{org}$ nor $test_{tar}$.
The MTLCF$\_0.5\_0.5$ achieves a better performance on both the $test_{org}$ and $test_{tar}$, and has achieved $5\%$  relative reduction of CER on $test_{org}$, which is compared with the initial model in Table \ref{Table1}.

\begin{table}[!htbp]
\centering
\caption{Performance of the three Training methods on SwitchBoard corpus and Fisher corpus}
\begin{tabular}{ c | c | c | c }
\hline
Method    &   $scale_{tar}$  &   $test_{org}$(CER) & $test_{tar}$(CER)\\ \hline

FT   & $300$h  & $0.900442$  & $0.289869$ \\

RT   & $300$h     & $0.151849$  & $0.342297$\\

MTLCF    & $300$h    & \textbf{0.111771}  & \textbf{0.288898}\\
\hline
FT   & $120$h  & $0.789013$  & \textbf{0.323052} \\

RT   & $120$h     & $0.205404$  & $0.378759$\\

MTLCF    & $120$h    & \textbf{0.121136}  & $0.324862$
\\\hline
\end{tabular}
\label{Table2}
\end{table}

For the above phenomena, we analyze the reason is that learning the new knowledge while not forgetting the old knowledge can not only effectively prevent the CF but also improve the performance on the original domain, which had been proven by the Stein몶s paradox.
And it can also prove our previous assumptions about the change of parameters.

On the other hand, we analyze the results of an unequal amount of training data.
As can be seen from the last three rows in Table \ref{Table2}, we can find that it obtained the results similar to the above experiment.
This fully proves that our proposed method is still applicable when the data scale differs greatly.

\subsubsection{Effect of hyperparametrices to MTLCF}

\begin{figure}[!htbp]

\centering

\subfigure[]{
\label{fig:figure4:a}
\begin{minipage}[t]{0.5\linewidth}
\centering
\includegraphics[width=1.6in]{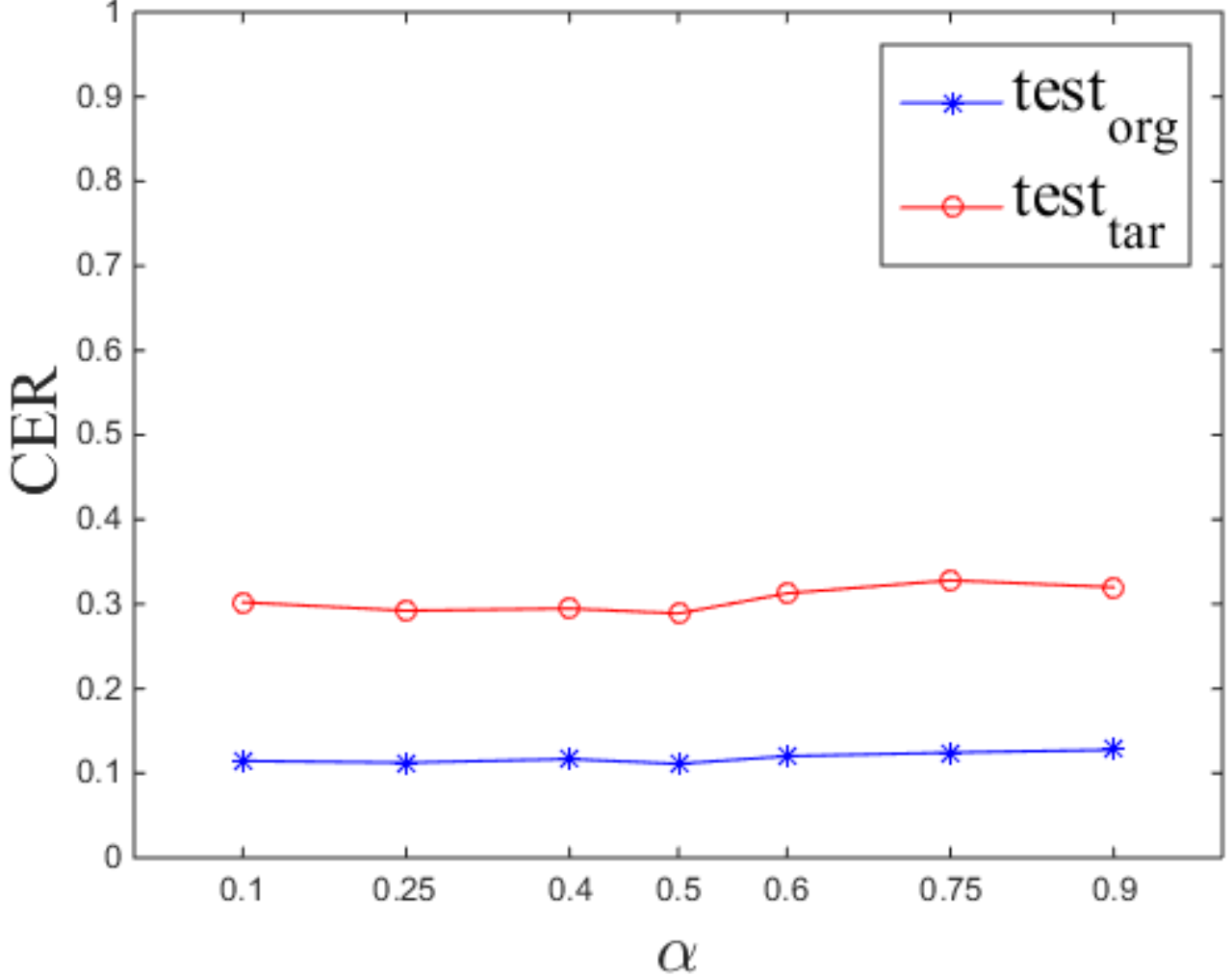}
\end{minipage}
}%
\subfigure[]{
\label{fig:figure4:b}
\begin{minipage}[t]{0.5\linewidth}
\centering
\includegraphics[width=1.6in]{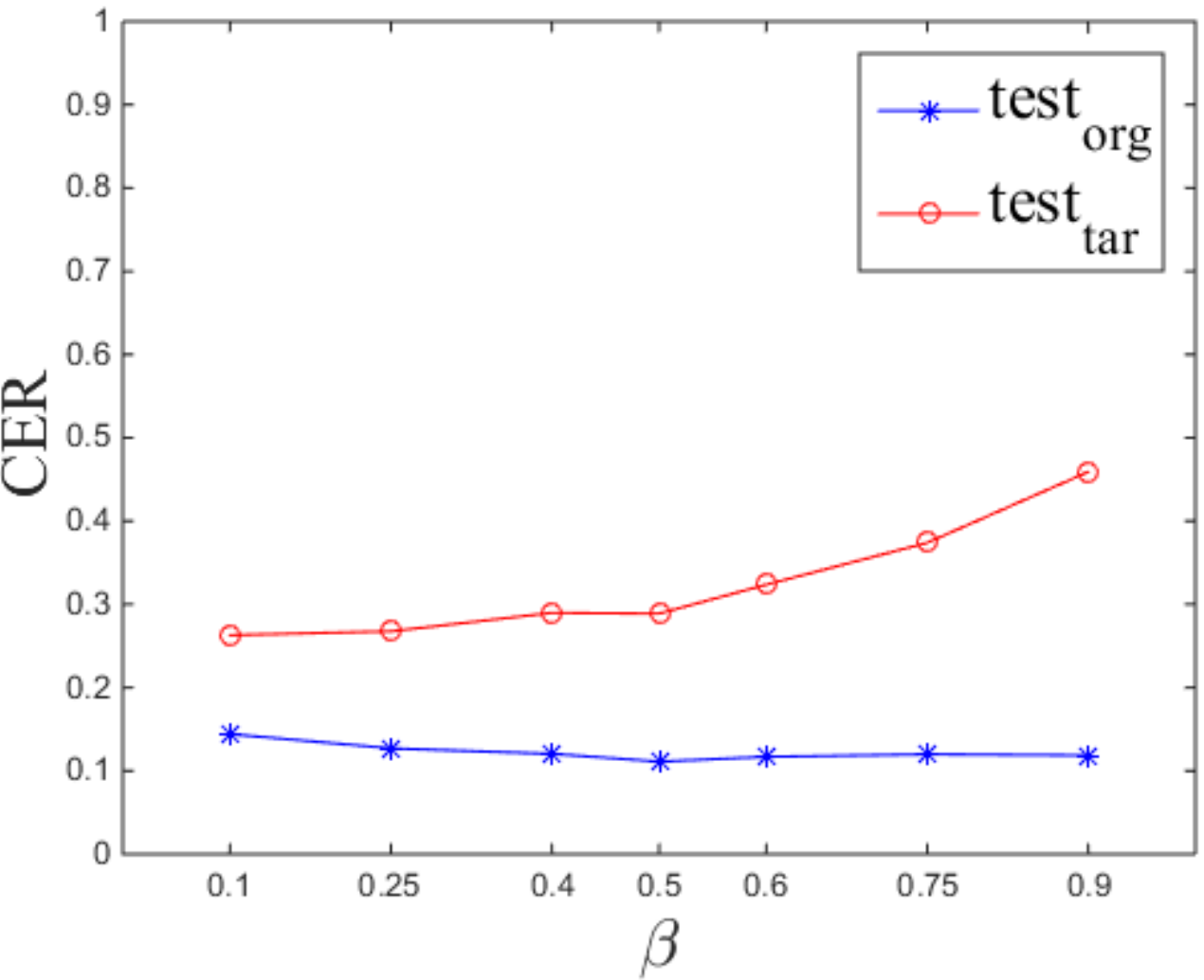}

\end{minipage}
}%

\centering
\caption{The influence of hyperparametric $\alpha$ on the convergence results of the new model when $\beta$ is 0.5 (a). The influence of hyperparametric $\beta$ on the convergence results of the new model when $\alpha$ is 0.5 (b).}
\label{Figure4}
\end{figure}

To further discuss the experimental results, we try to analyze the influence of different hyperparametrices on the training results of the MTLCF.
On the one hand, we analyze the influence of the change of $\alpha$ on the CER, which is shown in Figure \ref{Figure4}\subref{fig:figure4:a}.
In this figure, it can be found that the MTLCF model converges to a better results on both the $test_{org}$ and $test_{tar}$ when we choose $\alpha=0.5$.
And as $\alpha$ increases, the performance of the $test_{tar}$ continues to be degraded.
Therefore, in the following analysis we fixed $\alpha=0.5$.

On the other hand, the effect of the transformation of the hyperparametric $\beta$ on the model convergence results is shown in Figure \ref{Figure4}\subref{fig:figure4:b}.
From Figure \ref{Figure4}\subref{fig:figure4:b}, it can be easily seen that  with the increase of $\beta$, the proportion of $task1$ in the optimized gradient increases, this results in the lower error rate for $test_{org}$ and a higher error rate for  $test_{tar}$.
When $\beta=0.1$, the model achieved the best performance on the new dataset, and the CER dropped to $0.263$.

Through the above analysis, we find that the $\beta$ is more important to the convergence result of the MTLCF than $\alpha$, and we can adjust the performance of the original domain and the target domain by controlling the size of it.

\section{Conclusions}
Domain adaptation is an important topic for ASR systems.
In this paper, we attempt to overcome the CF in this process by using the lifelong learning, which adopts a new way of thinking about MTL.
We further propose a novel MTL based method (MTLCF) to learn new knowledge quickly without forgetting the learned knowledge.
We evaluate our proposed methods on the SwitchBoard corpus and LibriSpeech corpus.
From the experiment results, we can see that the proposed method achieves good performance on both the original domain and the target domain test set.
\section{Acknowledgements}
\label{ssec:Acknowledgements}

This research was supported by National Key Research and Development Plan of China under Grant 2017YFB1002102 and National Natural Science Foundation of China under Grant U1736210

\clearpage
\bibliographystyle{IEEEtran}

\bibliography{mybib}

\begin{thebibliography}{10}
\providecommand{\url}[1]{#1}
\csname url@samestyle\endcsname
\providecommand{\newblock}{\relax}
\providecommand{\bibinfo}[2]{#2}
\providecommand{\BIBentrySTDinterwordspacing}{\spaceskip=0pt\relax}
\providecommand{\BIBentryALTinterwordstretchfactor}{4}
\providecommand{\BIBentryALTinterwordspacing}{\spaceskip=\fontdimen2\font plus
\BIBentryALTinterwordstretchfactor\fontdimen3\font minus
  \fontdimen4\font\relax}
\providecommand{\BIBforeignlanguage}[2]{{%
\expandafter\ifx\csname l@#1\endcsname\relax
\typeout{** WARNING: IEEEtran.bst: No hyphenation pattern has been}%
\typeout{** loaded for the language `#1'. Using the pattern for}%
\typeout{** the default language instead.}%
\else
\language=\csname l@#1\endcsname
\fi
#2}}
\providecommand{\BIBdecl}{\relax}
\BIBdecl

\bibitem{sainath2015convolutional}
T.~N. Sainath, O.~Vinyals, A.~W. Senior, and H.~Sak, ``{Convolutional, Long
  Short-Term Memory, fully connected Deep Neural Networks},'' in \emph{{IEEE}
  International Conference on Acoustics, Speech and Signal Processing,
  {ICASSP}}, 2015, pp. 4580--4584.

\bibitem{DBLP:conf/icassp/ValtchevOWY96}
V.~Valtchev, J.~Odell, P.~C. Woodland, and S.~J. Young, ``Lattice-based
  discriminative training for large vocabulary speech recognition,'' in
  \emph{{IEEE} International Conference on Acoustics, Speech, and Signal
  Processing Conference Proceedings, {ICASSP}}, 1996, pp. 605--608.

\bibitem{amodei2016deep}
D.~Amodei, S.~Ananthanarayanan, R.~Anubhai, J.~Bai, E.~Battenberg, C.~Case,
  J.~Casper, B.~Catanzaro, Q.~Cheng, G.~Chen \emph{et~al.}, ``{Deep Speech 2 :
  End-to-End Speech Recognition in English and Mandarin},'' in
  \emph{International Conference on Machine Learning, {ICML}}, 2016, pp.
  173--182.

\bibitem{DBLP:conf/interspeech/PaulikFSSSW05}
M.~Paulik, C.~F{\"{u}}gen, S.~St{\"{u}}ker, T.~Schultz, T.~Schaaf, and
  A.~Waibel, ``Document driven machine translation enhanced {ASR},'' in
  \emph{{INTERSPEECH}}, 2005, pp. 2261--2264.

\bibitem{yu2010roles}
D.~Yu, L.~Deng, and G.~Dahl, ``Roles of pre-training and fine-tuning in
  context-dependent dbn-hmms for real-world speech recognition,'' in
  \emph{Proc. NIPS Workshop on Deep Learning and Unsupervised Feature
  Learning}, 2010.

\bibitem{french1999catastrophic}
R.~M. French, ``Catastrophic forgetting in connectionist networks,''
  \emph{Trends in cognitive sciences}, vol.~3, no.~4, pp. 128--135, 1999.

\bibitem{kirkpatrick2017overcoming}
J.~Kirkpatrick, R.~Pascanu, N.~Rabinowitz, J.~Veness, G.~Desjardins, A.~A.
  Rusu, K.~Milan, J.~Quan, T.~Ramalho, A.~Grabska-Barwinska \emph{et~al.},
  ``Overcoming catastrophic forgetting in neural networks,'' \emph{Proceedings
  of the national academy of sciences}, vol. 114, no.~13, pp. 3521--3526, 2017.

\bibitem{rusu2016progressive}
A.~A. Rusu, N.~C. Rabinowitz, G.~Desjardins, H.~Soyer, J.~Kirkpatrick,
  K.~Kavukcuoglu, R.~Pascanu, and R.~Hadsell, ``Progressive neural networks,''
  \emph{arXiv preprint arXiv:1606.04671}, 2016.

\bibitem{DBLP:conf/iclr/YoonYLH18}
J.~Yoon, E.~Yang, J.~Lee, and S.~J. Hwang, ``{Lifelong Learning with
  Dynamically Expandable Networks},'' in \emph{International Conference on
  Learning Representations, {ICLR}}, 2018.

\bibitem{stein1956inadmissibility}
C.~Stein, ``Inadmissibility of the usual estimator for the mean of a
  multivariate normal distribution,'' STANFORD UNIVERSITY STANFORD United
  States, Tech. Rep., 1956.

\bibitem{caruana1997multitask}
R.~Caruana, ``Multitask learning,'' \emph{Machine learning}, vol.~28, no.~1,
  pp. 41--75, 1997.

\bibitem{panayotov2015librispeech}
V.~Panayotov, G.~Chen, D.~Povey, and S.~Khudanpur, ``{Librispeech: An {ASR}
  corpus based on public domain audio books},'' in \emph{{IEEE} International
  Conference on Acoustics, Speech and Signal Processing, {ICASSP}}, 2015, pp.
  5206--5210.

\bibitem{godfrey1992switchboard}
J.~J. Godfrey, E.~C. Holliman, and J.~McDaniel, ``{SWITCHBOARD: Telephone
  speech corpus for research and development},'' in \emph{{IEEE} International
  Conference on Acoustics, Speech and Signal Processing, {ICASSP}}, 1992, pp.
  517--520.

\bibitem{hecht1992theory}
R.~Hecht-Nielsen, ``Theory of the backpropagation neural network,'' in
  \emph{Neural networks for perception}.\hskip 1em plus 0.5em minus 0.4em\relax
  Elsevier, 1992, pp. 65--93.

\bibitem{sussmann1992uniqueness}
H.~J. Sussmann, ``Uniqueness of the weights for minimal feedforward nets with a
  given input-output map,'' \emph{Neural networks}, vol.~5, no.~4, pp.
  589--593, 1992.

\bibitem{graves2006connectionist}
A.~Graves, S.~Fern{\'{a}}ndez, F.~J. Gomez, and J.~Schmidhuber, ``Connectionist
  temporal classification: labelling unsegmented sequence data with recurrent
  neural networks,'' in \emph{Machine Learning, Proceedings of the
  International Conference {ICML}}, 2006, pp. 369--376.

\bibitem{chua1988cellular}
L.~O. Chua and L.~Yang, ``Cellular neural networks: Theory,'' \emph{IEEE
  Transactions on circuits and systems}, vol.~35, no.~10, pp. 1257--1272, 1988.

\bibitem{hochreiter1997long}
S.~Hochreiter and J.~Schmidhuber, ``Long short-term memory,'' \emph{Neural
  Computation}, vol.~9, no.~8, pp. 1735--1780, 1997.

\bibitem{saon2017recent}
G.~Saon and M.~Picheny, ``Recent advances in conversational speech recognition
  using convolutional and recurrent neural networks,'' \emph{IBM Journal of
  Research and Development}, vol.~61, no.~4, pp. 1--1, 2017.

\bibitem{xiong2017microsoft}
W.~Xiong, J.~Droppo, X.~Huang, F.~Seide, M.~Seltzer, A.~Stolcke, D.~Yu, and
  G.~Zweig, ``{The Microsoft 2016 conversational speech recognition system},''
  in \emph{{IEEE} International Conference on Acoustics, Speech and Signal
  Processing, {ICASSP}}, 2017, pp. 5255--5259.

\bibitem{vesely2013sequence}
K.~Vesel{\`y}, A.~Ghoshal, L.~Burget, and D.~Povey, ``Sequence-discriminative
  training of deep neural networks.'' in \emph{{INTERSPEECH}}, 2013, pp.
  2345--2349.

\bibitem{povey2016purely}
D.~Povey, V.~Peddinti, D.~Galvez, P.~Ghahremani, V.~Manohar, X.~Na, Y.~Wang,
  and S.~Khudanpur, ``{Purely Sequence-Trained Neural Networks for ASR Based on
  Lattice-Free MMI.}'' in \emph{{INTERSPEECH}}, 2016, pp. 2751--2755.

\bibitem{medennikov2016improving}
W.~Hartmann, R.~Hsiao, T.~Ng, J.~Ma, F.~Keith, and M.-H. Siu, ``{Improved
  Single System Conversational Telephone Speech Recognition with VGG Bottleneck
  Features},'' in \emph{{INTERSPEECH}}, 2017, pp. 112--116.

\bibitem{povey2011kaldi}
D.~Povey, A.~Ghoshal, G.~Boulianne, L.~Burget, O.~Glembek, N.~Goel,
  M.~Hannemann, P.~Motlicek, Y.~Qian, P.~Schwarz \emph{et~al.}, ``The {Kaldi}
  speech recognition toolkit,'' in \emph{{IEEE} Workshop on Automatic Speech
  Recognition and Understanding, {ASRU}}, 2011, pp. 1--4.

\bibitem{pundak2016lower}
G.~Pundak and T.~N. Sainath, ``{Lower Frame Rate Neural Network Acoustic
  Models},'' in \emph{{INTERSPEECH}}, 2016, pp. 22--26.

\bibitem{abadi2016tensorflow}
M.~Abadi, P.~Barham, J.~Chen \emph{et~al.}, ``{TensorFlow: {A} System for
  Large-Scale Machine Learning},'' in \emph{12th {USENIX} Symposium on
  Operating Systems Design and Implementation, {OSDI}}, 2016, pp. 265--283.

\bibitem{audhkhasi2018building}
K.~Audhkhasi, B.~Kingsbury, B.~Ramabhadran, G.~Saon, and M.~Picheny,
  ``{Building Competitive Direct Acoustics-to-Word Models for English
  Conversational Speech Recognition},'' in \emph{{IEEE} International
  Conference on Acoustics, Speech and Signal Processing, {ICASSP}}, 2018, pp.
  4759--4763.

\bibitem{DBLP:journals/corr/KingmaB14}
D.~P. Kingma and J.~Ba, ``{Adam: {A} Method for Stochastic Optimization},'' in
  \emph{International Conference on Learning Representations, {ICLR}}, 2015.

\end{thebibliography}

\end{document}